# A Portable High-Quality Random Number Generator for Lattice Field Theory Simulations

Martin Lüscher

Deutsches Elektronen-Synchrotron DESY
Notkestrasse 85, D-22603 Hamburg, Germany

## Abstract

The theory underlying a proposed random number generator for numerical simulations in elementary particle physics and statistical mechanics is discussed. The generator is based on an algorithm introduced by Marsaglia and Zaman, with an important added feature leading to demonstrably good statistical properties. It can be implemented exactly on any computer complying with the IEEE–754 standard for single precision floating point arithmetic.

# 1. Introduction

Numerical simulations in elementary particle physics and statistical mechanics are increasingly performed on massively parallel computers. These machines offer unmatched computing power, thus making it possible to simulate larger systems and to achieve greater statistical precision. It is well-known that the random number generators employed in these computations can be a source of systematic error. In fact many of the popular generators used to date failed to give correct results in some recent simulations of the 2-dimensional Ising model [1,2]. While the Ising model is a rather special case, with unusual regularity, the lesson clearly is that random number generators should be chosen with care, especially when one aims for high-precision results.

The generator discussed in this paper derives from an algorithm originally proposed by Marsaglia and Zaman [3]. It has a very long period and excellent statistical properties on short and long time scales. The quality of the generator is established using some mathematical results on chaotic dynamical systems, the spectral test and a number of empirical tests.

The algorithm has been implemented on the APE-100, a parallel computer now intensively used in elementary particle physics (for a short description and guide to the literature see ref.[4]). One may also easily write a FORTRAN code for the generator, which will run correctly on any machine complying with the IEEE-754 standard for single precision floating point arithmetic.

The definition and basic properties of the Marsaglia-Zaman algorithm are reviewed in sect. 2. For appropriately chosen parameters the period of the generator can be proved to be very large [3]. Its statistical properties are however not as good as initially assumed. In particular, the generator fails in the classical gap test [5] and an unfavourable lattice structure in the distribution of random numbers in high dimensions has been discovered [6,7].

The important new observation made in this paper is that the Marsaglia-Zaman algorithm is closely related to a dynamical system, which is known to be chaotic in a strong sense (it is a so-called $K$-system [11]). One then infers that the correlations detected in the gap test, for example, are short ranged in time. A sequence of random numbers with much better statistical properties is therefore obtained by picking out elements of the original sequence at time intervals greater than the correlation time. All this is explained in sects. 3 and 4, and the quality of the so improved generator is evaluated in sect. 5. Implementation details and timing benchmarks for various machines are included for completeness.

## 2. Marsaglia-Zaman generator

The random number generator defined below is based on a so-called subtract-with-borrow algorithm [3]. For the particular choice of parameters specified in subsect. 2.4 the generator is known by the name of RCARRY [8].

*2.1 Definition*

Let $b$ be an arbitrary integer greater than 1, referred to as the base, and define $X$ to be the set of integers $x$ satisfying $0 \leq x < b$. The algorithm generates a random sequence $x_0, x_1, x_2, \ldots$ of elements of $X$ recursively, together with a sequence $c_0, c_1, c_2, \ldots$ of "carry bits". The latter take values 0 or 1 and are used internally, i.e. the interesting output of the algorithm are the numbers $x_n$, or rather $x_n/b$, if one requires random numbers uniformly distributed between 0 and 1.

The recursion involves two fixed lags, $r$ and $s$, which are assumed to satisfy $r > s \geq 1$. For $n \geq r$ one first computes the difference

$$\Delta_n = x_{n-s} - x_{n-r} - c_{n-1}, \tag{2.1}$$

and then determines $x_n$ and $c_n$ through

$$\begin{aligned} x_n &= \Delta_n, & c_n &= 0 \quad \text{if} \quad \Delta_n \geq 0, \\ x_n &= \Delta_n + b, & c_n &= 1 \quad \text{if} \quad \Delta_n < 0. \end{aligned} \tag{2.2}$$

It is trivial to verify that $x_n$ is contained in $X$ if $x_{n-s}$ and $x_{n-r}$ are and if $c_{n-1}$ is 0 or 1. The name "carry bit" for $c_n$ is now quite intuitive, since $c_n$ simply indicates whether a shift by the base $b$ was necessary when computing $x_n$.

To start the recursion, the first $r$ values $x_0, x_1, \ldots, x_{r-1}$ together with an initial carry bit $c_{r-1}$ must be provided. The configurations

$$x_0 = x_1 = \ldots = x_{r-1} = 0, \qquad c_{r-1} = 0, \tag{2.3}$$

$$x_0 = x_1 = \ldots = x_{r-1} = b - 1, \qquad c_{r-1} = 1, \tag{2.4}$$

should be avoided, because the algorithm yields uninteresting sequences of numbers in these cases. All other choices of initial values are admitted in the following and we shall then say that the generator has been properly initialized.

*2.2 Period of the generator*

For some values of the base $b$ and the lags $r, s$, the period of the sequence generated through eqs.(2.1),(2.2) can be determined rigorously. Define

$$m = b^r - b^s + 1 \tag{2.5}$$

and let $q$ be the smallest positive integer such that

$$b^q = 1 \bmod m. \tag{2.6}$$

The existence of $q$ is guaranteed since $m$ and $b$ are relatively prime. An important mathematical result of Marsaglia and Zaman now is [3]

**Theorem 2.1.** *If $m$ is a prime number, the period of the generator defined through eqs.(2.1),(2.2) is equal to $q$. More precisely, if the generator has been properly initialized, the following is true.*

*1. For all $n \geq r$ we have $x_{n+q} = x_n$.*

*2. Any number $p$, such that $x_{n+p} = x_n$ for more than $r$ successive values of $n$, is an integer multiple of $q$.*

It should be emphasized that the period is independent of the chosen initial values $x_0, x_1, \ldots, x_{r-1}$. Note that this particular string of numbers may not occur anywhere else in the sequence, i.e. in general the algorithm gets into a loop only after the first $r$ updates have been made.

Another comment is that the period of the generator must be expected to depend on the initial values, if $m$ is *not* prime. Such generators are not safe and should be avoided unless all periods can be shown to be large.

*2.3 Associated linear congruential generator*

The algorithm of Marsaglia and Zaman is closely related to the standard linear congruential generator with multiplier

$$a = m - (m-1)/b \tag{2.7}$$

and modulus $m$ [6]. Such generators have been studied vigorously in the past and we shall later rely on some of this theory when we discuss the statistical properties of the random number sequence produced by the Marsaglia-Zaman algorithm.

The linear congruential generator alluded to above operates on the set of all integers $y$ in the range $0 < y < m$. Starting from an initial value $y_0$, a sequence of random numbers $y_0, y_1, y_2, \ldots$ is obtained recursively through

$$y_n = a y_{n-1} \bmod m. \tag{2.8}$$

The multiplier $a$ satisfies
$$ab = 1 \bmod m \tag{2.9}$$

and the recursion is thus equivalent to

$$y_n = b y_{n+1} \bmod m. \tag{2.10}$$

It is not difficult to show that the period of the sequence is equal to $q$ if $m$ is prime.

The relation between this generator and the Marsaglia-Zaman generator is summarized by [6]

**Theorem 2.2.** *Let $(x_n)_{n \geq 0}$ be a sequence of random numbers generated through the Marsaglia-Zaman algorithm, with carry bits $(c_n)_{n \geq 0}$ and proper initial values. Then, for all $n \geq r$, the integers*

$$y_n = \sum_{k=0}^{r-1} x_{n-r+k} b^k - \sum_{k=0}^{s-1} x_{n-s+k} b^k + c_{n-1} \tag{2.11}$$

*are in the range $0 < y_n < m$. Moreover the relation*

$$b y_{n+1} - y_n = m x_n \tag{2.12}$$

*holds and the sequence $(y_n)_{n \geq r}$ is thus generated through the recursion (2.8).*

The theorem shows at once that the Marsaglia-Zaman algorithm is essentially a clever way to implement certain linear congruential generators with huge moduli. Manipulations of large integers are avoided by breaking them up into a vector of smaller numbers which are then processed one by one.

## 2.4 Choice of parameters

Most computers used for large scale numerical simulations have been designed to yield maximum performance for floating point operations. The parameters $b$, $r$ and $s$ should thus be chosen so as to be able to implement the generator using floating point arithmetic.

Single precision real numbers on computers complying with the IEEE-754 standard are represented by a string of 32 bits, with 23 bits reserved for the mantissa and the rest for the sign and exponent of the number. Signed integers of absolute magnitude up to $2^{24}$ can thus be dealt with exactly on such machines using floating point arithmetic. So if we choose

$$b = 2^{24}, \tag{2.13}$$

all elements of $X$ (and $b$ itself) will be computer representable numbers. As for the lags $r$ and $s$, we take

$$r = 24, \qquad s = 10, \tag{2.14}$$

a choice proposed by Marsaglia and Zaman [3] and recommended by James [8]. The difference $\Delta_n$ in the recursion (2.2) then is

$$\Delta_n = x_{n-10} - x_{n-24} - c_{n-1}, \tag{2.15}$$

and 24 integers $x_0, x_1, \ldots, x_{23}$ in the range $0 \leq x_k < 2^{24}$ plus a carry bit $c_{23}$ are required to initialize the generator †. Note that no rounding occurs in the computation of $\Delta_n$, since the final and intermediate results are representable numbers, i.e. the algorithm is implemented exactly.

The modulus $m$ and multiplier $a$ for this choice of parameters are given by

$$m = 2^{576} - 2^{240} + 1, \tag{2.16}$$

$$a = 2^{576} - 2^{552} - 2^{240} + 2^{216} + 1. \tag{2.17}$$

Using elementary number theory and the complete decomposition of $m - 1$ into prime factors, it is possible to prove that $m$ is a prime number [3]. The

---

† The FORTRAN code for this algorithm printed in ref.[8] contains an error. A correct program is obtained by interchanging the indices I24 and J24 in the line UNI=SEEDS(I24)-SEEDS(J24)-CARRY [9].

period of the generator is thus determined by theorem 2.1. Some further work then yields

$$q = (m-1)/48 \simeq 5.2 \times 10^{171}, \qquad (2.18)$$

which is a very long period indeed. There is no chance that, on any earthly computer, one will ever come close to exhausting this sequence of random numbers.

In the following the parameters of the generator are assumed to be as specified above. The reader should however meet no difficulty in carrying over the discussion to any other case of interest.

## 3. Origin of statistical correlations

The Marsaglia-Zaman generator is now known to fail in several empirical tests of randomness, a particularly simple case being the gap test ([5]; for a lucid description of the test see ref.[10], p.60f). As explained below there are in fact some rather obvious correlations between successive vectors of $r$ random numbers. They are seen most clearly when the generator is described in the language of dynamical systems.

### 3.1 Geometrical preliminaries

The unit hyper-cube in $r$ dimensions is the set of all vectors

$$v = (v_0, v_1, \ldots, v_{r-1}) \qquad (3.1)$$

with real components between 0 and 1. If opposite faces of the hyper-cube are identified one obtains an $r$ dimensional torus $T^r$. The points on this manifold are also represented by vectors $v$, as above, with the understanding that $v$ and $w$ describe the same point if $v_k = w_k \bmod 1$ for all $k$.

$T^r$ contains a discrete subset, $\dot{T}^r$, which consists of all vectors $v$ with components of the form

$$v_k = n_k/b, \qquad n_k = 0, 1, 2, \ldots, b-1. \qquad (3.2)$$

$\dot{T}^r$ is an $r$ dimensional hyper-cubic lattice with spacing $1/b$, which may be regarded as a discrete approximation of the torus.

The distance between any two points $v$ and $w$ on $T^r$ is defined through

$$d(v, w) = \max_k d_k, \qquad d_k = \min\{|v_k - w_k|, 1 - |v_k - w_k|\}. \qquad (3.3)$$

It is straightforward to check that $d$ has all the properties required for a decent distance function on $T^r$. In particular, it is invariant under translations modulo 1.

*3.2 The Marsaglia-Zaman generator as a dynamical system*

Let us now consider a sequence of random numbers $x_0, x_1, x_2, \ldots$ generated through the Marsaglia-Zaman algorithm, with carry bits $(c_n)_{n\geq 0}$ and proper initial values. The vectors

$$v(t) = (x_n, x_{n+1}, \ldots, x_{n+r-1})/b, \qquad n = rt, \qquad (3.4)$$

define a point on the (discrete) torus $\dot{T}^r$ which moves as the "time" $t$ progresses from 0 in steps of 1. If we also introduce a time dependent carry bit,

$$c(t) = c_{rt+r-1}, \qquad (3.5)$$

it is clear that the evolution of $v(t)$ and $c(t)$ is determined by the recursion (2.1),(2.2).

We are thus led to interpret the Marsaglia-Zaman generator as a discrete dynamical system, consisting of a set $S$ of states and a mapping $\phi : S \mapsto S$. A state is defined by a point on the discrete torus and a carry bit. $\phi$ maps any such state onto the next one, viz.

$$\bigl(v(t+1), c(t+1)\bigr) = \phi\bigl(v(t), c(t)\bigr). \qquad (3.6)$$

Note that $\phi$ does not refer to any of the previous states. One only needs to know the current state to be able to compute the next one.

*3.3 Continuity and statistical correlations*

For a good generator one requires that successive vectors of random numbers be statistically independent. That is, if $(v, c)$ runs through all possible states, the joint distribution of $(v, c)$ and $\phi(v, c)$ should be uniform on $S \times S$.

Of course this cannot be true since $\phi$ operates on a finite set of states. The distribution is at best approximately uniform. Since one can only generate a relatively small number of states in practice, one is anyway unable to test the distribution very precisely. One should however be worried by correlations that are strong enough to give a measurable effect in any simple statistical test.

We now show that such correlations exist. Let us first ignore the carry bits. The recursion (2.1),(2.2) then reads

$$x_n = x_{n-s} - x_{n-r} \mod b \tag{3.7}$$

and $\phi$ becomes a linear transformation of the torus. An important consequence of this fact is that nearby points are mapped onto nearby ones. So if one chooses a set of random points $v$ in some small volume, their successors $\phi(v)$ are contained in some other small volume. In particular, they are not scattered over the whole torus, as one would expect if $\phi(v)$ were statistically independent of $v$.

The carry bits only affect the least significant digits of the random numbers and so cannot destroy the basic continuity of $\phi$. More precisely, if we define

$$(\hat{v}, \hat{c}) = \phi(v, c), \tag{3.8}$$

it is possible to show that

$$d(\hat{v}, \hat{w}) \leq 4d(v, w) + 3/b. \tag{3.9}$$

The distance between two points on $\dot{T}^r$ thus increases by at most a factor 4 plus 3 lattice spacings. In particular, small regions are mapped onto small regions and so we again conclude that successive vectors of random numbers are strongly correlated.

It should be emphasized that the effects caused by these correlations are readily seen in empirical tests. In particular, the failure of the Marsaglia-Zaman generator in the gap test can be explained in this way. Note, incidentally, that similar correlations are present in all lagged Fibonacci generators using addition or subtraction as the binary operation.

# 4. Deterministic chaos

A characteristic feature of chaotic dynamical systems is that trajectories starting at nearby states diverge exponentially with time. Even if the evolution is locally continuous, such a system appears to behave randomly on larger time scales. One could also say that any state specified to some finite precision has an exponentially deteriorating memory of its history. We now show that the dynamical system underlying the Marsaglia-Zaman generator is chaotic in this sense.

*4.1 Numerical experiment*

It is helpful to start with a simple experiment illustrating the chaotic nature of the mapping $\phi$. The experiment consists in choosing a random sample of 1000 pairs of trajectories $(v(t), c(t))$ and $(v'(t), c'(t))$, with initial values separated by 1 lattice spacing, viz.

$$d(v(0), v'(0)) = 1/b. \tag{4.1}$$

One then computes the average distance

$$\delta(t) = \langle d(v(t), v'(t)) \rangle \tag{4.2}$$

as a function of the evolution time $t$.

Fig. 1 shows that the trajectories are rapidly diverging. In the range $4 \leq t \leq 16$ the data are well described by

$$\delta(t) = A e^t, \qquad A = 5 \times 10^{-8}, \tag{4.3}$$

i.e. the separation is growing exponentially with a rate close to 1. Around $t = 17$, $\delta(t)$ levels off and assumes a value equal to $12/25$ within statistical errors. This is the average distance between two randomly chosen points on the torus, thus indicating that $v(t)$ and $v'(t)$ are no longer correlated.

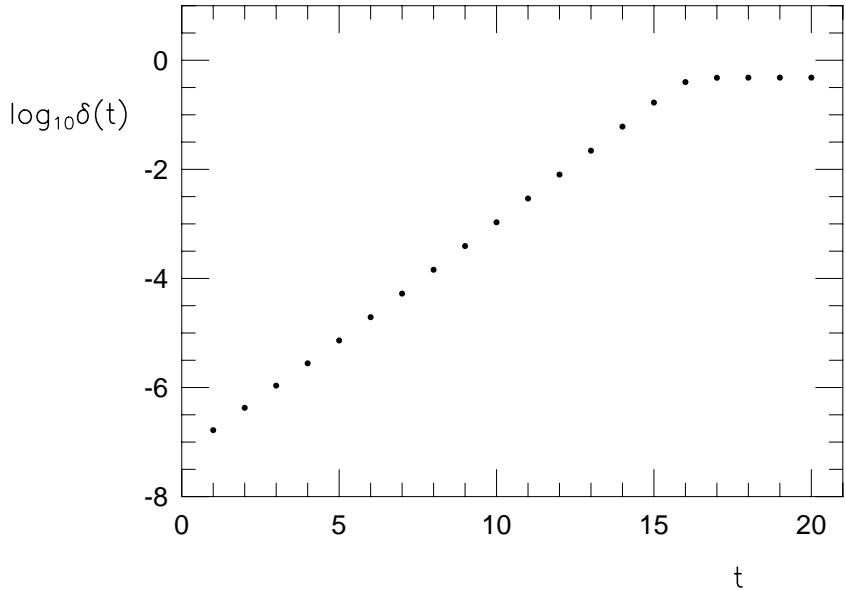

Fig. 1. Average distance $\delta(t)$ between neighbouring trajectories as a function of the evolution time $t$.

*4.2 Continuum limit*

For the further study of deterministic chaos it is now useful to pass to the continuum limit $1/b \to 0$, where the space of states $S$ becomes equal to the full torus $T^r$ and the carry bit is neglected. This is an accurate approximation to the discrete system on short time scales and if all distances of interest are much greater than the lattice spacing. In particular, the evolution of diverging trajectories can be expected to be correctly described when they are sufficiently far apart.

In the continuum limit the mapping $\phi$ reduces to

$$\phi(v) = L^r v \bmod 1, \qquad (4.4)$$

where $L$ is the linear transformation

$$Lv = (v_1, v_2, \ldots, v_{r-1}, v_{r-s} - v_0). \qquad (4.5)$$

$L$ can be considered an $r \times r$ matrix with entries $0, 1$ and $-1$. It is then trivial to verify that $\det L = 1$ and $\phi$ is hence invertible and volume preserving.

According to the established mathematical terminology, the continuum system $(T^r, \mu, \phi)$ (where $\mu$ denotes the standard measure on $T^r$) is a classical dynamical system. The occurence of chaos in such systems has been studied extensively and many deep results have been obtained. In the rest of this section the system $(T^r, \mu, \phi)$ will be discussed from the point of view of the mathematical theory. Although no previous knowledge on dynamical systems is required, the reader may now find it useful to consult one or the other book on the subject such as refs.[11–13], for example.

*4.3 Liapunov exponent*

In the continuum system the exponential rate of divergence of neighbouring trajectories can be computed analytically as follows.

Suppose $v(t)$ and $v'(t)$ are two trajectories such that their distance is very much smaller than 1 at $t = 0$. Let us define the difference vector

$$u(t) = v'(t) - v(t) \bmod 1, \qquad -\tfrac{1}{2} < u_k(t) \leq \tfrac{1}{2}. \tag{4.6}$$

It is clear that the norm of this vector,

$$\|u(t)\| = \max_k |u_k(t)|, \tag{4.7}$$

is equal to the distance between the trajectories at time $t$. Furthermore, from eq.(4.4) one infers that

$$u(t+1) = L^r u(t) \tag{4.8}$$

if $\|L^r u(t)\| < \tfrac{1}{2}$, a condition which is satisfied as long as the trajectories are sufficiently close.

The dominant exponential growth of the deviation vector $u(t)$ is hence determined by the largest eigenvalues of $L$. The characteristic equation of $L$,

$$\lambda^r - \lambda^{r-s} + 1 = 0, \tag{4.9}$$

can easily be solved numerically and one finds that all eigenvalues $\lambda$ are complex and non-degenerate. There are 4 eigenvalues with maximal absolute value given by

$$|\lambda|_{\max} = 1.04299\ldots \tag{4.10}$$

Now if the initial deviation vector $u(0)$ has a non-zero component in the direction of the corresponding eigenvectors (which is the generic case), one concludes that

$$\|u(t)\| \propto e^{\nu t} \qquad (4.11)$$

at large times $t$, where

$$\nu = r \ln |\lambda|_{\max} = 1.01027\ldots \qquad (4.12)$$

Of course eq.(4.11) only holds as long as the evolution equation (4.8) applies. By considering smaller and smaller initial deviations, this condition will be fulfilled for any desired length of time. Eq.(4.11) then becomes asymptotically exact.

The exponent $\nu$ is referred to as the Liapunov exponent of the system. As already noted in subsect. 4.2, the evolution of diverging trajectories in the discrete system is expected to be accurately described by the continuum system. A comparison of the result of the experiment, eq.(4.3), with the value of the Liapunov exponent $\nu$ confirms this. We have thus shown that the chaotic behaviour of the Marsaglia-Zaman generator can be traced back to the instability of the underlying lagged Fibonacci generator.

*4.4 Kolmogorov entropy and mixing*

The continuum system $(T^r, \mu, \phi)$ can be proved to belong to a class of strongly unstable systems. While the relevance of this remark for the discrete system is not completely obvious, it does provide some further insight into how repeated application of a smooth mapping can lead to randomness.

The mapping $\phi$ is in many respects similar to the famous cat map of Arnold. In particular, under the action of $\phi$ the torus is stretched in $r/2$ directions and shrunk in $r/2$ complementary directions. After many iterations any region in $T^r$ (a cat's body, for example) is first made very long and thin and then wrapped on the torus. As a result the region is scattered over the whole manifold.

These heuristic remarks can be made much more precise and it is then possible to show, using the theorems discussed in ref.[11], that $(T^r, \mu, \phi)$ is a so-called $K$-system. This means that it has a positive Kolmogorov entropy and that consequently it is mixing and ergodic.

The property of mixing is particularly intuitive. It states that

$$\lim_{t \to \infty} \mu \left( A \cap \phi^t(B) \right) = \mu(A) \mu(B) \qquad (4.13)$$

for all measurable sets $A, B$. In other words, if the set $B$ is evolved for a long time, it will be uniformly distributed over the torus and thus occupies a fraction $\mu(B)$ of every other set $A$ (recall that $\phi$ is volume preserving).

The Kolmogorov entropy is a substantially more difficult notion. Basically it is the rate at which the knowledge about the system is lost as it evolves from an only imprecisely specified initial state. A positive entropy thus implies that one loses information exponentially fast.

## 5. Improved generator

The important qualitative implication of the chaotic nature of $\phi$ is that the correlations discovered in sect. 3 are short ranged in time. A sequence of random numbers with significantly better statistical properties is therefore obtained by keeping only a fraction of the full sequence of numbers produced by the Marsaglia-Zaman algorithm. The precise rule is given below and several statistical tests are performed to confirm the expected improvement.

### 5.1 Definition

We again start from a sequence of random numbers $x_0, x_1, x_2, \ldots$ generated through the Marsaglia-Zaman algorithm, with carry bits $(c_n)_{n \geq 0}$ and proper initial values. Instead of using all numbers $x_n$, we now read $r$ successive elements of the sequence, discard the next $p - r$ numbers, read $r$ numbers, and so on. The integer $p \geq r$ is a fixed parameter which allows us to monitor the fraction of random numbers "thrown away". In particular, the old generator corresponds to $p = r$, where no numbers are discarded.

The numbers selected in this manner define a history of states $\bigl(v(t), c(t)\bigr)$ through
$$v(t) = (x_n, x_{n+1}, \ldots, x_{n+r-1})/b, \qquad n = pt, \tag{5.1}$$
$$c(t) = c_{n+r-1}.$$

As before the time evolution is generated by a well-defined mapping $\phi_p : S \mapsto S$ such that
$$\bigl(v(t+1), c(t+1)\bigr) = \phi_p\bigl(v(t), c(t)\bigr). \tag{5.2}$$

In the continuum limit $\phi_p$ reduces to the linear transformation
$$\phi_p(v) = L^p v \bmod 1, \tag{5.3}$$

where $L$ is given by eq.(4.5).

The discussion in sect. 4 now suggests that deterministic chaos leads to a complete decorrelation of successive states for values of $p$ greater than about $16r = 384$. For such $p$ the corresponding sequence of random numbers is expected to possess excellent statistical properties. In practice one may be satisfied with a smaller value of $p$, as a full decorrelation, down to the level of the least significant bits, may in many cases be unnecessary. The statistical tests reported in the following subsections help to clarify the situation and a more definite recommendation as to which value of $p$ to choose will be issued after that.

*5.2 Spectral test*

For any state $(v,c)$ an integer $y$ in the range $0 < y < m$ may be defined through

$$y = \sum_{k=0}^{r-1} v_k b^{k+1} - \sum_{k=0}^{s-1} v_{r-s+k} b^{k+1} + c, \tag{5.4}$$

where $v_0, v_1, \ldots, v_{r-1}$ are the components of $v$ (cf. theorem 2.2). $y$ should be regarded as an observable constructed from the given state. In particular, a trajectory $\bigl(v(t), c(t)\bigr)$ of states, generated by the mapping $\phi_p$, is associated with a sequence of values $y(t)$. Theorem 2.2 tells us that

$$y(t+1) = a^p y(t) \bmod m, \tag{5.5}$$

i.e. $\phi_p$ is related to a linear congruential generator with modulus $m$ and multiplier $a^p \bmod m$.

The multi-dimensional distributions of $y(t)$ can be studied by applying the powerful spectral test for linear congruential generators. The test effectively probes the statistical independence of successive states $\bigl(v(t), c(t)\bigr)$, since any correlation between the values of $y(t)$ can be regarded as a correlation among the corresponding states. For a detailed description of the spectral test the reader is referred to Knuth's book [10]. Here we merely introduce the necessary notations and discuss the results of the test.

An infamous property of linear congruential generators is that vectors of $D$ successive random numbers fall on parallel hyper-planes with often appreciable spacing. The spectral test consists in calculating the maximal spacing $h_D$, or rather the "accuracy" $\nu_D = 1/h_D$, for low dimensionalities $D$. The

Table 1. Merits $\mu_D$ of some generators with modulus $m$ and multiplier $a^p \bmod m$

| $p$ | $\mu_2$ | $\mu_3$ | $\mu_4$ | $\mu_5$ | $\mu_6$ | $\mu_7$ | $\mu_8$ |
|---|---|---|---|---|---|---|---|
| 24 | $4\epsilon^{29}$ | $\epsilon^{85}$ | $\epsilon^{56}$ | $2\epsilon^{27}$ | $\epsilon^{86}$ | $3\epsilon^{72}$ | $6\epsilon^{58}$ |
| 48 | 0.20 | 0.07 | 0.03 | $9\epsilon^{23}$ | 5.08 | $2\epsilon^{33}$ | $2\epsilon^{31}$ |
| 96 | 2.67 | 1.04 | 1.64 | 0.04 | 1.60 | 0.14 | 0.10 |
| 192 | 1.82 | 0.67 | 0.70 | 1.53 | 2.69 | 4.78 | 1.54 |
| 384 | 0.56 | 0.82 | 2.30 | 1.56 | 0.84 | 4.60 | 0.29 |
| 768 | 1.63 | 2.59 | 3.08 | 0.59 | 0.96 | 1.29 | 1.12 |
| 223 | 1.80 | 0.87 | 2.39 | 3.79 | 2.29 | 0.78 | 2.29 |
| 389 | 2.27 | 3.46 | 3.92 | 2.49 | 2.98 | 4.23 | 0.46 |

[$\epsilon = \frac{1}{10}$; $m$ and $a$ are given by eqs.(2.16),(2.17)]

outcome of the spectral test may be rated through the figures of merit

$$\mu_D = \frac{(\nu_D \sqrt{\pi})^D}{m\Gamma\left(\frac{1}{2}D + 1\right)}. \tag{5.6}$$

Good generators achieve values of $\mu_D$ greater than 1 for say $D = 2, \ldots, 6$. On the other hand, if the merit is significantly smaller than 0.1 for some of these dimensions, one has picked a particularly bad multiplier.

The results of the spectral test are listed in table 1. The first line corresponds to the original generator where no random numbers are discarded. As already noted in refs.[6,7], there are strong correlations between successive values of the observable $y$ in this case, for any dimensionality $D$. Evidently this generator is a poor source of random numbers.

In general the merits are quite acceptable for $p$ greater than about 200. The merits for two favoured values around 200 and 400 are listed in the last two lines of table 2. All this is very much in line with what one expects from deterministic chaos. It should however be emphasized that the spectral test is a full period test, while the decorrelation through diverging trajectories takes place on short time scales.

### 5.3 Further statistical tests

a. *Serial correlation test.* This test is applied to the associated linear congruential generator. It is a full-period theoretical test, where one computes the correlation coefficient between successive values of $y$ exactly (see ref.[10] for further explanations). For values of $p$ greater than about 100 it is passed easily.

b. *Gap test.* In ref.[5] the original generator ($p = 24$) has been subjected to a large number of empirical tests. All tests were passed with the exception of the gap test. This test has now been repeated for various values of $p$, with the same test parameters, and no significant statistical correlations were detected for $p \geq 48$.

c. *Ising model.* Simulations of the 2-dimensional Ising model, using cluster algorithms, have proved to be a particularly sensitive test of random number generators [1,2]. Such a test has recently been performed by Wolff [14] for $p = 223$ and $p = 389$. In both cases no discrepancy between the simulation data and the exact analytic results was found.

d. *SU(2) lattice gauge theory.* The generator with $p = 223$ is now being used in some high-precision calculations of the running coupling in the SU(2) lattice gauge theory [15]. So far all results obtained are compatible with earlier computations where shift register generators were employed.

### 5.4 Recommended values of p

From the theoretical discussion and the tests of the improved generator one concludes that the remaining statistical correlations are small when $p$ is greater than about 200. The recommended default value is $p = 223$, and if one has any doubts that the simulation results might be biased by the random number generator, one may still set $p = 389$. A decorrelation of successive vectors of $r$ random numbers down to the least significant digits is then guaranteed.

To take still larger values of $p$ appears to be pointless, since no empirical test or theoretical consideration indicates that a further improvement will be achieved.

Table 2.  Average time needed to produce 1 new random number ($p = 223$)

| machine | time [$\mu$s] |
|---|---|
| SUN 10-41 | 5 |
| HP 9000/735 | 2 |
| CRAY YMP (1 CPU) | 0.7 |
| APE-100 (1 node) | 5 |

*5.5 Implementation and timing*

As discussed in sect. 2, the Marsaglia-Zaman algorithm can be implemented exactly using single precision floating point arithmetic. If random numbers between 0 and 1 are desired, it is advantageous to work directly with the numbers $x_n/b$ instead of $x_n$. No rounding is implied by this renormalization since $b$ is a power of 2, i.e. the implementation remains exact.

A portable FORTRAN code for the improved generator has been developed by James [16] and is available through the CPC library. The name of the program is RANLUX. It comes with an initialization subroutine and further entry points to save and read the state of the generator.

The generator has also been implemented on the APE-100 parallel computer [17]. The program may be obtained through anonymous ftp by dialing 141.108.16.27 and copying the contents of the directory pub/random, or by writing to the author (luscher@ips102.desy.de).

Since one uses only a fraction of the basic sequence of random numbers, the improved generator tends to be slow. For numerical simulations of lattice field theories, where large quantitites of random numbers are requested, it is hence important to take full advantage of any pipelining capabilities of the hardware. A difficulty here is that the Marsaglia-Zaman recursion (2.1),(2.2) refers to the carry bit $c_{n-1}$ computed in the preceding step and so is not suitable for vectorization.

The problem can be overcome by running several copies of the generator in parallel, with different initial values. The arithmetic operations are then pipelined horizontally, i.e. when looping over the copies. On the APE-100, for example, a good efficiency is achieved with 24 copies on each node. Some care should of course be paid to properly initialize the generators. In view of the astronomical period of the generator, the chances that any two of the copies yield significantly correlated random numbers are however extremely slim.

Some timing benchmarks for the improved generator with $p = 223$ are listed in table 2 [14,17]. The programs were written in FORTRAN and APESE, a high-level language for the APE-100. It is obvious that the numbers quoted depend on many technical details. They should hence be interpreted as a rough estimate of what can be achieved with a modest programming effort.

## 6. Concluding remarks

A well-known problem with random number generators is that their quality is difficult to assess in any rigorous way. Some confidence in the reliability of any given generator can of course be gained by performing a large number of statistical tests. But doubts will always remain that the generator might fail in the next test.

There exists an impressive list of classical dynamical systems which have been shown to be strongly chaotic. The states in these systems move randomly on time scales substantially greater than a certain characteristic time, related to the Liapunov exponent of the system. It should be emphasized that randomness can be given a precise mathematical meaning in this framework.

The random number generator discussed in this paper may be considered a discrete approximation to such a chaotic dynamical system. A theoretical understanding of why the algorithm yields statistically independent random numbers is thus obtained. On longer time scales theoretical support for the good quality of the generator comes from the spectral test and the fact that the period can be shown to be extremely long. One might object that the generator is too slow for large scale applications. But other parts of the program are often much more costly so that the extra computer time needed for the generator is insignificant. One may also prefer to pay the price rather than taking any risk of producing corrupted data, especially when spending months of parallel computer time to a single project.

I would like to thank Ulli Wolff for performing the Ising model tests and providing some of the timing benchmarks quoted in table 2. I am also indebted to Fred James for various useful informations and constant encouragement. Helpful discussions with Kari Kankaala, Rainer Sommer, Marcus Speh, Frank Steiner and Peter Weisz are gratefully acknowledged.

# References

[1] A. M. Ferrenberg, D.P. Landau, and Y. J. Wong, Phys. Rev. Lett. 69 (1992) 3382

[2] P. D. Coddington, Analysis of Random Number Generators Using Monte Carlo Simulation, preprint, Northeast Parallel Architectures Center, Syracuse University (1993)

[3] G. Marsaglia and A. Zaman, Ann. Appl. Prob. 1 (1991) 462

[4] E. Marinari, Nucl. Phys. B (Proc. Suppl.) 30 (1993) 122

[5] I. Vattulainen, K. Kankaala, J. Saarinen and T. Ala-Nissila, A Comparative Study of Some Pseudorandom Number Generators, preprint, University of Helsinki HU-TFT-93-22, hep-lat 9304008

[6] S. Tezuka, P. L'Ecuyer and R. Couture, On the Lattice Structure of the Add-With-Carry and Subtract-With-Borrow Random Number Generators, preprint (1993)

[7] R. Couture and P. L'Ecuyer, On the Lattice Structure of Certain Linear Congruential Sequences Related to AWC/SWB Generators, preprint (1993)

[8] F. James, Comp. Phys. Commun. 60 (1990) 329

[9] F. James, private communication (1993)

[10] D. E. Knuth, Semi-Numerical Algorithms, *in*: The Art of Computer Programming, vol. 2, 2nd ed. (Addison-Wesley, Reading MA, 1981)

[11] V. I. Arnold and A. Avez, Ergodic Problems of Classical Mechanics (Addison-Wesley, Redwood City, 1989)

[12] H. G. Schuster, Deterministic Chaos, 2nd ed. (VCH Verlagsgesellschaft, Weinheim, 1989)

[13] A. M. Ozorio de Almeida, Hamiltonian Systems: Chaos and Quantization (Cambridge University Press, Cambridge, 1988)

[14] U. Wolff, private communication (1993)

[15] R. Frezzotti, M. Guagnelli, M. Lüscher, R. Petronzio, R. Sommer, P. Weisz and U. Wolff, work in progress

[16] F. James, Comp. Phys. Commun., to appear

[17] M. Lüscher, A Random Number Generator for the APE-100 Parallel Computer, unpublished internal report (June 1993)